\documentclass[12pt]{iopart}

\begin{document}

\title{Generalized Kaluza-Klein monopole, quadratic algebras and ladder operators}

\author{Ian Marquette}

\address{Department of Mathematics, University of York, Heslington, York, UK. YO10 5DD}
\ead{im553@york.ac.uk}
\begin{abstract}
We present a generalized Kaluza-Klein monopole system. We solve this quantum superintegrable systems on a Euclidean Taub Nut manifold using the separation of variables of the corresponding Schroedinger equation in spherical and parabolic coordinates. We present the integrals of motion of this system, the quadratic algebra generated by these integrals, the realization in term of a deformed oscillator algebra using the Daskaloyannis construction and the energy spectrum. The structure constants and the Casimir operator are functions not only of the Hamiltonian but also of other two integrals commuting with all generators of the quadratic algebra and forming an Abelian subalgebra. We present an other algebraic derivation of the energy spectrum of this system using the factorization method and ladder operators.
\end{abstract}

\maketitle

\section{Introduction}

The idea of monopole in context of quantum mechanics was discussed first by Dirac [1]. However, the problem of degeneracy in the presence of a magnetic monopole was studied later, independently by Zwanziger [2] and by McIntosh and Cisneros [3]. This `MICZ-Kepler' that consist of a magnetic monopole in flat 3D space with an arbitrary electric charge plus a fixed inverse-{\em square} term in the potential, is superintegrable (i.e. it allows more integrals of motion than degrees of freedom), and has an $so(4)$ symmetry algebra generated by the Poincar\'e vector together with a Runge-Lenz vector. This problem was also discussed by Barut et al.[4], Jackiw [5] and including spin by D'Hoker and Vinet [6]. The MICZ-Kepler is separable in spherical and parabolic coordinates, this multiseparability is related with the existence of these integrals of motion. The MICZ-Kepler problem also exists in higher dimensions [7-14], and remain superintegrable. The Yang-Coulomb system, consisting  of the Yang monopole and a particle coupled to the monopole by isospin $su(2)$ and the Kepler-Coulomb interaction [7-12]. This system was studied using the separation of variables in hyperspherical, spheroidal and parabolic coordinates and the interbasis expansion obtained [10]. This system is also related to the 8D harmonic oscillator by a Hurwitz transformation and has a $so(6)$ symmetry algebra. The MICZ-Kepler problem was also considered in $S^{3}$ [15]. 

A generalized MICZ-Kepler was introduced by Mardoyan [16]. This system can be seen as the intrinsic Smorodinsky-Winternitz system [17-19] with monopole in 3D Euclidean space. It can be transformed into a 4D singular harmonic oscillator by a duality transformation [20]. This 4D singular harmonic oscillator studied in [21] and the classical analog and its periodic trajectories in [22]. This 4D system allows also to discuss Poschl-Teller, Hartmann potential and other ring shaped system in a unified way. We presented the quadratic algebra of the generalized MICZ-Kepler system in three-dimensional Euclidean space $E_{3}$ and its dual the four dimensional singular oscillator in four-dimensional Euclidean space $E_{4}$ [23]. We presented their realization in terms of a deformed oscillator algebra using the Daskaloyannis construction [24] and presented an algebraic derivation of the energy spectrum. We also presented a new algebraic derivation of the energy spectrum of the MICZ-Kepler system on the three sphere $S^{3}$ [15] using a quadratic algebra. These results pointed out also that results and explicit formula for structure functions obtained for quadratic, cubic and higher order polynomial algebras in context of two-dimensional superintegrable systems without monopole may be applied to superintegrable systems in higher dimensions with and without monopoles. The generalized MICZ-Kepler system [25] were discussed from the point of view of supersymmetric quantum mechanics and ladder operators [26]. In context of superintegrable systems without monopole such relation between superintegrability, ladder operators and supersymmetric quantum mechanics [27-34]. Method to obtain new superintegrable systems were proposed from ladder and supersymmetry in quantum mechanics.  

Let us introduce an other class of monopole system. The idea of five-dimensional theory with one dimensional curled up to form a circle was discussed in context of unification theory by Kaluza and Klein [35,36]. Later, the Taub Nut metric (with a cyclic coordinates) appeared in context of the Euclidean Einstein equation [37], the Taub Nut metric is Ricci-flat self dual on $R^{4}$ and gives an example of non trivial gravitational instanton. A Kaluza-Klein monopole was obtained by embedding the Taub-Nut gravitational instanton into five-dimensional Kaluza-Klein theory [38,39] and also appeared in context of the study of monopole scattering [40-42]. Since, many article were devoted to Kaluza-Klein monopole [43-47]. It was observed than in classical mechanics the equation of motion contains the Dirac monopole, a Coulomb term and a velocity-square dependent term, thus this system appears to be a non trivial generalization of the Kepler system. More interestingly, the Kaluza-Klein monopole in classical and quantum mechanics possesses conserved quantities that are analog of the angular momentum and the Runge-Lenz-vector and an algebraic derivation of it discrete energy spectrum can be done from it dynamical symmetry algebra o(4)[44,45]. This systems is multiseparable and the solution of the Schrödinger equation in spherical and parabolic coordinates [46,47] was done and the interbasis expansion obtained [46]. Moreover, analog of such systems can be obtained for generalized Taub Nut metric [48-51] and Lebrun metric [52 53]. Recently, intrinsic Smorodinsky-Winternitz systems on various curved space such the Taub Nut metric in a classical context using the $sl(2,R)$ coalgebra symmetry approach were introduced [54].

The purpose of this paper is to study a quantum superintegrable generalization of the Kaluza-Klein monopole that correspond to an intrinsic Smorodinsky-Winternitz systems obtain its integrals of motion and the corresponding quadratic algebra, present an algebraic derivation of the energy spectrum using this quadratic algebra and also using ladder operators.

The paper is structured as follows. In Section 2, recall some results of the Kaluza-Klein monopole and we present a generalization of Kaluza-Klein monopole and solve the corresponding Schrodinger equation in spherical and parabolic coordinates. In Section 3, recall results on quadratic algebras and their realization in term of deformed oscillator algebras and present for this system the quadratic algebra, the Casimir operators, the realization in terms of a deformed oscillator algebra and its spectrum. In Section 4, using factorization and ladder operators we reobtain the energy spectrum.

\section{Kaluza-Klein monopole}

Let us recall some results of the Kaluza-Klein monopole. The Euclidean Taub Nut metric is given by 
\begin{equation}
ds^{2}-dt^{2}+\frac{1}{V(r)}dl^{2}+V(r)(dx^{5}+A_{i}dx^{i})^{2},
\end{equation}

with

\begin{equation}
V(r)=\frac{1}{1+\frac{\mu}{r}},\quad A_{1}=\frac{\mu}{r}\frac{x_{2}}{r+x_{3}},\quad A_{2}=\frac{\mu}{r}\frac{x_{1}}{r+x_{3}},\quad A_{3}=0,
\end{equation}

where $dl^{2}=(dx^{1})^{2}+(dx^{2})^{2}+(dx^{3})^{2}$ is the 3-dimensional Euclidean line element.

The Hamiltonian for the quantum Kaluza-Klein monopole is given by :

\begin{equation}
H=\frac{1}{2}(V(r) \vec{P}^{2}+\frac{1}{V(r)}P_{5}^{2}),\quad 
\end{equation}

where the operators

\begin{equation}
P_{5}=-i\partial_{5},\quad P_{i}=-i(\partial_{i}-A_{i}\partial_{5}),
\end{equation}

satisfy the following commutation relations :

\begin{equation}
[P_{i},P_{j}]=i\epsilon_{ijk}B_{k}P_{5},\quad [P_{i},P_{5}]=0.
\end{equation}

where $\vec{B}=\mu \frac{\vec{x}}{r^{3}}$. Let us now consider the change of variables $x^{5}=\mu\psi$ and define the charge operator $Q=-i\mu\partial_{\psi}$. The Hamiltonian allows the following first and second order integrals of motion that consist in the angular momentum and Runge-Lenz vector:

\begin{equation}
\vec{L}=\vec{x}\times \vec{P}-\frac{\vec{x}}{r}Q, \quad \vec{K}=\frac{1}{2}(\vec{P}\times \vec{L}-\vec{L}\times \vec{P})-\mu\frac{\vec{x}}{r}(H-\frac{1}{\mu^{2}}Q^{2}).
\end{equation}

These integrals generate a so(4) dynamical symmetry algebra :

\begin{equation}
[L_{i},L_{j}]=i\epsilon_{ijk}L_{k},\quad [L_{i},K_{j}]=i\epsilon_{ijk}K_{k},\quad [K_{i},K_{j}]=i\epsilon_{ijk}L_{k}(\frac{Q^{2}}{\mu^{2}}-2H).
\end{equation}

Similarly to the Kepler-Coulomb systems the energy spectrum can be obtained algebraically from the finite-dimensional unitary representations and is given by the following expression ($Q\Psi=q\Psi$, $L_{3}\Psi=m\Psi$ and $L^{2}\Psi=l(l+1)\Psi$) :

\begin{equation}
n=\frac{(\frac{q^{2}}{\mu}-\mu E)}{\sqrt{\frac{q^{2}}{\mu^{2}}-2E}},\quad n=n_{r}+l+1,\quad n_{r}=0,1,2,...
\end{equation}

with the following constraints $|q|-1<l \leq n-1$ and $|q|-1<|m|\leq l$.

\subsection{Generalized Kaluza-Klein monopole}

Let us now introduce the following Hamiltonian with $V(r)$ and the components $A_{i}$ given by Eq(2):

\begin{equation}
H=\frac{1}{2}(V(r)(\vec{P}^{2}+\frac{c_{1}}{2r}+\frac{c_{2}}{2r(r+z)}+\frac{c_{3}}{2r(r-z)}+c_{4})+\frac{1}{V(r)}P_{5}^{2}).
\end{equation}

Let us postpone the study of the integrals of motion and the symmetry algebra of this Hamiltonian in the next section and solve the corresponding Schroedinger equation in spheroidal and parabolic coordinates systems.

Considering the spheroidal coordinates systems 
\begin{equation}
x=rsin(\theta)cos(\phi),\quad y=rsin(\theta)sin(\phi),\quad z=rcos(\theta),
\end{equation}

the the Taub Nut metric thus has the form :

\begin{equation}
ds^{2}=-dt^{2}+\frac{1}{V}(dr^{2}+r^{2}d\theta^{2}+r^{2}sin^{2}(\theta)d\phi^{2})+\mu^{2}(d\psi+(1-cos(\theta))d\phi)^{2},
\end{equation}
\begin{equation}
A_{r}=A_{\theta}=0,\quad A_{\phi}=\mu(1-cos(\theta)),
\end{equation}

and the Schroedinger equation takes the following form :

\begin{equation}
-\frac{1}{2}V(r)(\frac{1}{r^{2}}\partial_{r}(r^{2}\partial_{r})-\frac{c_{1}}{2r}-\frac{c_{2}}{4r^{2}cos^{2}(\frac{\theta}{2})}-\frac{c_{3}}{4r^{2}sin^{2}(\frac{\theta}{2})}-c_{4}
\end{equation}
\[+\frac{1}{r^{2}}(\frac{\partial^{2}}{\partial \theta^{2}}+cot(\theta)\frac{\partial}{\partial \theta}+\frac{1}{sin^{2}(\theta)}\frac{\partial^{2}}{\partial \phi^{2}})+(\frac{1}{\mu^{2}V(r)^{2}}+\frac{(1-cos(\theta))^{2}}{r^{2}sin^{2}(\theta)})\frac{\partial^{2}}{\partial \psi^{2}}\]
\[\frac{-2}{r^{2}}\frac{(1-cos(\theta))}{sin^{2}(\theta)}\frac{\partial}{\partial \phi}\frac{\partial}{\partial \psi})\Psi(r,\theta,\phi,\psi)=E\Psi(r,\theta,\phi,\psi).\]

By making the Ansatz :

\begin{equation}
\Psi(r,\theta,\phi,\psi)=R(r)Z(\theta)e^{i(\nu\phi+q\psi)},
\end{equation}

we obtain for the angular and radial variables the following ordinary differential equations :

\begin{equation}
\frac{1}{r^{2}}\frac{d}{dr}(r^{2}\frac{dR}{dr})+(  (2E -\frac{q^{2}}{\mu^{2}}-c_{4})+\frac{1}{r}(2E\mu -\frac{2q^{2}}{\mu}-c_{1})-\frac{A}{r^{2}})R=0,
\end{equation}

\begin{equation}
\frac{\partial^{2}Z}{\partial \theta^{2}}+cot(\theta)\frac{\partial Z}{\partial \theta}+(A-\frac{(m-q)^{2}+c_{2}}{4cos^{2}(\frac{\theta}{2})}-\frac{(m+q)^{2}+c_{3}}{4sin^{2}(\frac{\theta}{2})})Z=0.
\end{equation}

where $L_{3}\Psi(r,\theta,\phi,\psi)=m\Psi(r,\theta,\phi,\psi)$, $Q \Psi(r,\theta,\phi,\psi)=q \Psi(r,\theta,\phi,\psi)$ ($\nu=m-q$).

The solution of Eq(2.16) is given by 

$Z=N_{jm}(\delta_{1},\delta_{2})(cos(\frac{\theta}{2}))^{m_{1}} (sin(\frac{\theta}{2}))^{m_{2}}P_{j-m_{+}}^{m_{2},m_{1}}(cos(\theta))$, 

with $m_{1}=|m-q|+\delta_{1}=\sqrt{(m-q)^{2}+c_{2}}$, $m_{2}=|m+s|+\delta_{2}=\sqrt{(m+q)^{2}+c_{3}}$, $m_{+}=\frac{(|m+q|+|m-q|)}{2}$, $P_{n}^{a,b}$ denotes a Jacobi polynomial. The quantum number $m$ and $j$ run through values $m=-j,-j+1,...,j-1,j$ and $j=m_{+},m_{+}+1$,.... . The quantum numbers j,m characterise the total momentum of the system and its projection on the z axis. A is a separation constant quantized as $A=(j+\frac{\delta_{1}+ \delta_{2}}{2})(j+ \frac{\delta_{1}+ \delta_{2}}{2}+1)$. 

For the radial variable, we introduce $R=\frac{1}{r}\chi_{E,j}^{q}(r)$ and obtain the differential equation :

\begin{equation}
(-\frac{d^{2}}{dr^{2}}+\frac{A}{r^{2}}-\frac{\alpha}{r})\chi_{E,j}^{q}(r)=\beta \chi_{E,j}^{q}(r),
\end{equation}
with
\begin{equation}
\alpha=2E\mu -\frac{2q^{2}}{\mu}-\frac{c_{1}}{2},\quad \beta=2E-\frac{q^{2}}{\mu^{2}}-c_{4}.
\end{equation}

The solution is given in term of the confluent hypergeometric function
\begin{equation}
\chi(r)=c_{nj}(\delta_{1},\delta_{2})\rho^{s}e^{-\frac{\rho}{2}}F(a,2s,\rho),
\end{equation}

where $c$ is a normalization constant with
\begin{equation}
\rho=2r \sqrt{-\beta},\quad a=s-\frac{\alpha}{2\sqrt{-\beta}}.
\end{equation}

We obtain from $s=j+\frac{\delta_{1}+\delta_{2}}{2}$ and $a=-n_{r}$ ($n_{r}=0,1,2,...$) the following condition :
\begin{equation}
j+\frac{\delta_{1}+\delta_{2}}{2}-\frac{\alpha}{2\sqrt{-\beta}}=-n_{r},
\end{equation}

that gives the energy spectrum

\begin{equation}
\frac{(2E\mu-\frac{2q^{2}}{\mu}-\frac{c_{1}}{2})}{2\sqrt{c_{4}-2E+\frac{q^{2}}{\mu^{2}}}}=n+\frac{\delta_{1}+\delta_{2}}{2},\quad n=n_{r}+j+1.
\end{equation}

In term of the parabolic coordinates :

\begin{equation}
x_{1}=\sqrt{\xi \eta}cos(\phi),\quad x_{2}=\sqrt{\xi \eta}sin(\phi),\quad z=\frac{1}{2}(\xi-\eta),\quad r=\frac{1}{2}(\xi+\eta),
\end{equation}

The the Taub Nut metric take the following form :

\begin{equation}
ds^{2}=-dt^{2}+\frac{1}{V(r)}dl^{2}+\mu^{2}V(r)(d\psi +(1-\frac{\xi-\eta}{\xi+\eta})d\phi)^{2},
\end{equation}
\begin{equation}
dl^{2}=\frac{\xi+\eta}{4\xi}d\xi^{2}+\frac{\xi+\eta}{4\eta}d\eta^{2}+\xi \eta d\phi^{2}.
\end{equation}

The Hamiltonian has the form :

\begin{equation}
H=\frac{1}{2}(V(\vec{P}^{2}+\frac{c_{1}}{\xi+\eta}+\frac{c_{2}}{(\xi+\eta)\xi}+\frac{c_{3}}{(\xi+\eta)\eta}+c_{4})+\frac{1}{V}\frac{Q^{2}}{\mu^{2}}),
\end{equation}

and can be written in term of $L_{3}$ and $Q$ as :

\begin{equation}
H=\frac{1}{2\mu^{2}}Q^{2}-\frac{2}{\xi+\eta+2\mu}(\partial_{\xi}(\xi\partial_{\xi})+\partial_{\eta}(\eta\partial_{\eta})-\frac{c_{1}}{4}-\frac{c_{2}}{4\xi}-\frac{c_{3}}{4\eta}-\frac{c_{4}(\xi+\eta)}{4}
\end{equation}
\[-\frac{1}{4\xi}(L_{3}-Q)^{2}-\frac{1}{4\eta}(L_{3}+Q)^{2}-\frac{1}{2}Q^{2}).\]

By making the following Ansatz :

\begin{equation}
\Psi(\xi,\eta,\phi,\psi)=f_{E,k,m}^{q}(\xi)h_{E,k,m}^{q}(\eta)e^{i(\nu\phi+q\psi)},
\end{equation}

and using the method of separation of variables we obtain the following set of ordinary differential equations :

\begin{equation}
(\partial_{\xi}(\xi \partial_{\xi})-\frac{m_{1}^{2}}{4\xi}+\frac{\alpha}{4}+\frac{\beta}{4}\xi)f_{E,k,m}^{q}(\xi)=\frac{k}{2}f_{E,k,m}^{q}(\xi),
\end{equation}
\begin{equation}
(\partial_{\eta}(\eta \partial_{\eta})-\frac{m_{2}^{2}}{4\eta}+\frac{\alpha}{4}+\frac{\beta}{4}\eta)h_{E,k,m}^{q}(\eta)=-\frac{k}{2}h_{E,k,m}^{q}(\eta).
\end{equation}

The solutions of the Eq(29) and (30) is given in term of the confluent hypergeometric function :

\begin{equation}
f_{E,k,m}^{q}(\xi)=x^{\frac{m_{1}}{2}}e^{-\frac{x}{2}}F(-n_{1},m_{1}+1,x),\quad x=\xi\sqrt{-\beta},
\end{equation}

\begin{equation}
h_{E,k,m}^{q}(\eta)=y^{\frac{m_{2}}{2}}e^{-\frac{y}{2}}F(-n_{2},m_{2}+1,y),\quad y=\eta\sqrt{-\beta},
\end{equation}

where $n_{1}$ and $n_{2}$ are non negative integers and

\begin{equation}
n_{1}=-\frac{m_{1}+1}{2}+\frac{(2k-\alpha)}{4\sqrt{-\beta}},\quad n_{2}=-\frac{m_{2}+1}{2}-\frac{(2k+\alpha)}{4\sqrt{-\beta}}.
\end{equation}

The principal quantum number $n$ can be expressed in term of the parabolic quantum numbers $n_{1}$ and $n_{2}$ :

\begin{equation}
n=n_{1}+n_{2}+m_{+}+1,\quad n_{1}, n_{2}=0,1,2,...\quad .
\end{equation}

\section{Quadratic algebra and realizations in term of deformed oscillator algebras}

The most general quadratic algebra [24] generated by the integrals of motion of a quadratically superintegrable system is given by :
\[ [A,B]=C, \]
\begin{equation}
[A,C]=\beta A^{2} +\gamma\{A,B\}+\delta A+\epsilon B+\zeta,
\end{equation}
\[ [B,C]=aA^{2}-\gamma B^{2}-\beta \{A,B\}+d A-\delta B + z.\]

the structure constants $\beta$, $\delta$, $\gamma$, $\epsilon$, $\zeta$, $a$, $d$ and $z$ can be polynomial functions not only of the Hamiltonian but also of any other integrals of motion ($F_{i}$) that commute with the Hamiltonian, with each other and also with the generators of the quadratic algebra $A$, $B$ and $C$ (that is, $[F_{i},H]=[F_{i},F_{j}]=[F_{i},A]=[F_{i},B]=[F_{i},C]=0$). Let us mention that the study of finite-dimensional unitary representations of a general cubic algebra [27] and example of quintic and seventh order algebras [30] were done. 

When we study this algebra's realization in terms of deformed oscillator algebras and its representations, we will fixed the energy ($H\psi=E\psi$) but also these other integrals ($F_{i}\psi=f_{i}\psi$) of motion that form, with the Hamiltonian, an Abelian subalgebra. The Casimir operator ($[K,A]=[K,B]=[K,C]=0$) of this quadratic algebra is thus given in terms of the generators by :
\begin{equation*}
K=C^{2}-\alpha\{A^{2},B\}-\gamma\{A,B^{2}\}+(\alpha \gamma -\delta)\{A,B\}+(\gamma^{2}-\epsilon)B^{2}
\end{equation*}
\[+(\gamma \delta -2 \zeta)B+\frac{2a}{3}A^{3}+(d+\frac{a\gamma}{3}+\alpha^{2})A^{2}+(\frac{a\epsilon}{3}+\alpha\delta+2z)A.\]

The  Casimir operator $K$ will be also rewritten as a polynomial of $H$ and $F_{i}$. 

A deformed oscillator algebra $\{N,b^{\dagger},b\}$ satisfy the following relations :
\begin{equation}
[N,b^{\dagger}]=b^{\dagger},\quad [N,b]=-b,\quad bb^{\dagger}=\Phi(N+1),\quad b^{\dagger}b=\Phi(N),
\end{equation}
where $\Phi(x)$ the structure function , is a real function. Two types of realizations of the quadratic algebra in terms of deformed oscillator algebras were obtained in Ref.24.

\textbf{Case} $\gamma \neq 0$ :
\begin{equation*}
\rho(N)=\frac{1}{2^{12}3\gamma^{8}(N+u)(1+N+u)(1+2(N+u))^{2}},\quad A(N)=\frac{\gamma}{2}((N+u)-\frac{1}{4}-\frac{\epsilon}{\gamma^{2}},
\end{equation*}
\begin{equation*}
b(N)=-\frac{\alpha((N+u)^{2}-\frac{1}{4})}{4}+\frac{\alpha\epsilon-\delta\gamma}{2\gamma^{2}}-\frac{\alpha\epsilon^{2}-2\delta\gamma\epsilon+4\gamma^{2}\zeta}{4\gamma^{4}}\frac{1}{((N+u)^{2}-\frac{1}{4})},
\end{equation*}
\begin{equation}
\Phi(N)=-3072\gamma^{6}K(-1+2(N+u))^{2}\label{eq12}
\end{equation}
\[-48\gamma^{6}(\alpha^{2}\epsilon-\alpha\delta\gamma+a\epsilon\gamma-d\gamma^{2})(-3+2(N+u))(-1+2(N+u))^{2}(1+2(N+u))\]
\[+\gamma^{8}(3\alpha^{2}+4a\gamma(-3+2(N+u))^{2}(-1+2(N+u))^{4}(1+2(N+u))^{2}\]
\[+768(\alpha\epsilon^{2}-2\delta\epsilon\gamma+4\gamma^{2}\zeta)^{2}+32\gamma^{4}(-1+2(N+u))^{2}(-1-12(N+u)\]
\[+12(N+u)^{2})(3\alpha^{2}\epsilon^{2}-6\alpha\delta\epsilon\gamma+2a\epsilon^{2}\gamma+2\delta^{2}\gamma^{2}-4d\epsilon\gamma^{2}+8\gamma^{3}z+4\alpha\gamma^{2}\zeta)\]
\[-256\gamma^{2}(-1+2(N+u))^{2}(3\alpha^{2}\epsilon^{3}-9\alpha\delta\epsilon^{2}\gamma+a\epsilon^{3}\gamma+6\delta^{2}\epsilon\gamma^{2}-3d\epsilon^{2}\gamma^{2}\]
\[+2\delta^{2}\gamma^{4}+2d\epsilon\gamma^{4}+12\epsilon\gamma^{3}z-4\gamma^{5}z+12\alpha\epsilon\gamma^{2}\zeta-12\delta\gamma^{3}\zeta+4\alpha\gamma^{4}\zeta).\]

\textbf{Case} $\gamma=0$, $\epsilon \neq 0$ :
\begin{equation*}
A(N)=\sqrt{\epsilon}(N+u),\quad b(N)=-\alpha(N+u)^{2}-\frac{\delta}{\sqrt{\epsilon}}(N+u)-\frac{\zeta}{\epsilon},\quad \rho(N)=1,
\end{equation*}
\begin{equation}
\Phi(N)=\frac{1}{4}(-\frac{K}{\epsilon}-\frac{z}{\sqrt{\epsilon}}-\frac{\delta}{\sqrt{\epsilon}}\frac{\zeta}{\epsilon}+\frac{\zeta^{2}}{\epsilon^{2}})\label{eq13}
\end{equation}
\[-\frac{1}{12}(3d-a\sqrt{\epsilon}-3\alpha\frac{\delta}{\sqrt{\epsilon}}+3(\frac{\delta}{\sqrt{\epsilon}})^{2}-6\frac{z}{\sqrt{\epsilon}}+6\alpha\frac{\zeta}{\epsilon}-6\frac{\delta}{\sqrt{\epsilon}}\frac{\zeta}{\epsilon})(N+u)\]
\[+\frac{1}{4}(\alpha^{2}+d-a\sqrt{\epsilon}-3\alpha\frac{\delta}{\sqrt{\epsilon}}+(\frac{\delta}{\sqrt{\epsilon}})^{2}+2\alpha\frac{\zeta}{\epsilon})(N+u)^{2}\]
\[-\frac{1}{6}(3\alpha^{2}-a\sqrt{\epsilon}-3\alpha\frac{\delta}{\sqrt{\epsilon}})(N+u)^{3}+\frac{1}{4}\alpha^{2}(N+u)^{4}.\]
 
To obtain finite-dimensional unitary representations [24] we should impose the following three constraints on the structure function:
\begin{equation}
\Phi(p+1,u,k)=0, \quad \Phi(0,u,k)=0,\quad \phi(x)>0, \quad \forall \; x>0 \quad .
\end{equation}
where $p=0,1,2,...$. The energy $E$ and the arbitrary constant $u$ are solutions of the equation obtained from these constraints. The dimension of the representation given by Eq.(39) is $p+1$.

\subsection{Quadratic algebra and generalized Kaluza-Klein monopole}

The Hamiltonian given by Eq.(9) allows the following integrals of motion :

\begin{equation}
A=L_{1}^{2}+L_{2}^{2}+L_{3}^{2}+\frac{c_{2}}{4}\frac{\eta}{\xi} +\frac{c_{3}}{4}\frac{\xi}{\eta},
\end{equation}
\begin{equation}
B=\frac{1}{2}((P_{1}L_{2}-P_{2}L_{1})-(L_{1}P_{2}-L_{2}P_{1}))-\mu\frac{(\xi-\eta)}{(\xi+\eta)}(H-\frac{1}{\mu^{2}}Q^{2}-\frac{c_{1}}{4\mu}+\frac{c_{2}\eta^{2}-c_{3}\xi^{2}}{2\mu\xi\eta(\xi-\eta)}),
\end{equation}
\begin{equation}
L_{3}=x_{1}P_{2}-x_{2}P_{1}-\frac{(\xi-\eta)}{(\xi+\eta)}Q,\quad Q=-i\mu\partial_{\psi}.
\end{equation}

In addition to these commutation relations :

\begin{equation}
[H,Q]=[H,L_{3}]=[H,A]=[H,B]=0,
\end{equation}

we have the further commutation relations :

\begin{equation}
[L_{3},Q]=[A,Q]=[B,Q]=[A,L_{3}]=[B,L_{3}]=0.
\end{equation}

Thus the integrals $H$, $Q$ and $L_{3}$ form an Abelian subalgebra.

We can form for the integrals $A$ and $B$ respectively related with the separation of variables in spherical and parabolic coordinates the following quadratic algebra :
\begin{equation}
[A,B]=C,
\end{equation}
\[ [A,C]=2\{A,B\}-4\mu HQL_{3}+\frac{4}{\mu}Q^{3}L_{3}+c_{1}QL_{3}+(c_{2}+c_{3})B+(c_{2}-c_{3})\mu H\]
\[+(\frac{c_{3}-c_{2}}{\mu})Q^{2}-\frac{1}{4}c_{1}(c_{2}-c_{3}),\]
\[ [B,C]=-2B^{2}+8AH+2\mu^{2}H^{2}-8HQ^{2}-4HL_{3}^{2}-\frac{4}{\mu^{2}}AQ^{2}+\frac{4}{\mu^{2}}Q^{4}\]
\[+\frac{2}{\mu^{2}}Q^{2}L_{3}^{2}-4c_{4}A+(4-c_{1}\mu)H+(\frac{-2+c_{1}\mu+2c_{4}\mu^{2}}{\mu^{2}})Q^{2}+2c_{4}L_{3}^{2}+\frac{1}{8}(c_{1}^{2}-16c_{4}).\]

\begin{equation}
K=4\mu^{2}H^{2}L_{3}+4\mu^{2}H^{2}Q^{2}+\mu^{2}(c_{2}+c_{3})H^{2}-8HQ^{4}-16HQ^{2}L_{3}^{2}-2c_{1}\mu HQ^{2}
\end{equation}
\[+2(c_{2}+c_{3}-c_{1}\mu)HL_{3}^{2}+(-2c_{2}-2c_{3}+2c_{2}c_{3}-\frac{c_{1}c_{2}\mu}{2}-\frac{c_{1}c_{3}\mu}{2})H+4(c_{2}-c_{3})HQL_{3}\]
\[+\frac{4}{\mu^{2}}Q^{6}+\frac{8}{\mu^{2}}Q^{4}L_{3}^{2}+\frac{2c_{1}}{\mu}Q^{4}+(\frac{-c_{2}-c_{3}+2c_{1}\mu+4c_{4}\mu^{2}}{\mu^{2}})Q^{2}L_{3}^{2}\]
\[-2\frac{(c_{2}-c_{3})}{\mu^{2}}L_{3}Q^{3}+(\frac{4(c_{2}+c_{3}-c_{2}c_{3})+2\mu c_{1}(c_{2}+c_{3})+4\mu^{2}(c_{1}^{2}-c_{2}c_{4}-c_{3}c_{4})}{4\mu^{2}})Q^{2}\]
\[-2c_{4}(c_{2}-c_{3})QL_{3}+(\frac{c_{1}^{4}}{4}-c_{2}c_{4}-c_{3}c_{4})L_{3}^{2}+\frac{1}{16}(c_{1}^{2}(c_{2}+c_{3})+16c_{4}(c_{2}+c_{3}-c_{2}c_{3})).\]

The structure constants of the quadratic algebra and the Casimir operator are thus written in term of the integrals forming the Abelian subalgebra as observed for other systems with and without monopole [23,55].

The structure function is given by the following equation :
\begin{equation}
\Phi(N)=-\frac{1}{\mu^{2}}12288(c_{2}^{2}-2c_{2}(c_{3}+(1-2(N+u))^{2}+4mq)+(c_{3}+(1+2m-2(N+u))
\end{equation}
\[(-1+2(N+u)+2q))(c_{3}-(-1+2m+2(N+u))(-1+2(N+u)-2q)))\]
\[(-16q^{4}+4q^{2}((1-2(N+u))^{2}-2c_{1}\mu + 8E\mu^{2})\]
\[-\mu^{2}(4(-c_{4}+2E)(1-2(N+u))^{2}+(c_{1}-4E\mu)^{2})).\]

The structure function can be factorized as :
\begin{equation}
\Phi(N)=\frac{2^{20}3}{\mu^{2}}(q-c_{4}\mu^{2}+2E\mu^{2})(N+u-\frac{1}{2}(1-(m_{1}-m_{2})))(N+u-\frac{1}{2}(1+(m_{1}-m_{2})))
\end{equation}
\[ (N+u-\frac{1}{2}(1-(m_{1}+m_{2})))(N+u-\frac{1}{2}(1+(m_{1}+m_{2})))]\]
\[ (N+u-(\frac{1}{2}-\frac{(4q^{2}+\mu(c_{1}-4E\mu))}{4\sqrt{q^{2}+(c_{4}-2E)\mu^{2}}})) (N+u-(\frac{1}{2}+\frac{(4q^{2}+\mu(c_{1}-4E\mu))}{4\sqrt{q^{2}+(c_{4}-2E)\mu^{2}}})).\]

From the Eq.(39) we obtain :

\begin{equation}
u=\frac{1}{2}+\frac{(4q^{2}+\mu(c_{1}-4E\mu))}{4\sqrt{q^{2}+(c_{4}-2E)\mu^{2}}},
\end{equation}

\begin{equation}
\Phi(N)=\frac{2^{20}3}{\mu^{2}}(-q-c_{4}\mu^{2}+2E\mu^{2})(p+1+m_{1}+m_{2}-N)N(p+1-N)
\end{equation}
\[(p+1+m_{2}-N)(p+1+m_{1}-N)(N-2p-2-m_{1}-m_{2}),\]

\begin{equation}
p+1+m_{1}+m_{2}=-\frac{(4q^{2}+\mu(c_{1}-4E\mu))}{4\sqrt{q^{2}+(c_{4}-2E)\mu^{2}}}.
\end{equation}

By taking taking $p=n_{1}+n_{2}$ and using the Eq.(35) we obtain that this energy spectrum coincide with results obtained from the separation of variables in Section 2.

\section{Generalized Kaluza-Klein monopole and ladder operators}

Let us now present a study of this systems using the factorization method and ladder operators. The radial part of the Schrodinger equation in spherical coordinates given by Eq.(17) using the change of variables $r= \frac{1}{\sqrt{-\beta}}x$ can be transformed in the following equation :

\begin{equation}
L_{n,j}\chi_{n,j}\equiv (-x^{2}\frac{d^{2}}{dx^{2}}-x^{2}-f_{n}x)\chi_{n,j}=-A\chi_{n,j},\quad f_{n}=\frac{\alpha}{\sqrt{-\beta}}.
\end{equation}

It can be factorized similarly to the generalized MICZ-Kepler system [26]. We can introduced the following operators :

\begin{equation}
B_{\pm}^{n}=\mp x\frac{d}{dx}+x-f_{n},
\end{equation}

satisfying the commutation relations :

\begin{equation}
(B_{-}^{n}-1)B_{+}^{n}\chi_{nj}=(f_{n}(f_{n}+1)-A)\chi_{nj},
\end{equation}
\begin{equation}
(B_{+}^{n}+1)B_{-}^{n}\chi_{nj}=(f_{n}(f_{n}-1)-A)\chi_{nj}.
\end{equation}

Let us define the following operators :

\begin{equation}
B_{3}=\frac{1}{2}(-x\frac{d^{2}}{dx^{2}}+x+\frac{A}{x}),\quad B_{\pm}=\mp x\frac{d}{dx}+x-B_{3}.
\end{equation}

They satisfy a su(1,1) dynamical Lie algebra as for the generalized MICZ-Kepler system [26] and various other quantum mechanics systems [56,57] and relativistic Kepler-coulomb problem [58] :

\begin{equation}
[B_{\pm},B_{3}]=\mp B_{\pm},\quad [B_{+},B_{-}]=-2B_{3}
\end{equation}

The Casimir operator is given by

\begin{equation}
B^{2}= -T_{\pm}T_{\mp}+T_{3}^{2}\mp T_{3}
\end{equation}

the Casimir $T^{2}$ and the operators $B_{3}$ act on the eigenfunctions in the following way

\begin{equation}
B_{3}\chi_{nj}=f_{n}\chi_{nj},\quad T^{2}\chi_{nj}=A\chi_{nj}.
\end{equation}

The unitary representations of $su(1,1)$ are given by

\begin{equation}
C^{2}\psi_{\nu,\mu}=\mu(\mu+1)\psi_{\mu,\nu},\quad C_{3}\psi_{\nu,\mu}=\nu \psi_{\nu,\mu},
\end{equation}

with $\nu=\mu+n'+1$,n'=0,1,2,... and $\mu>-1$. We have $\mu=j+\frac{\delta_{1}+\delta_{2}}{2}$ and we obtain

\begin{equation}
\nu=j+\frac{\delta_{1}+\delta_{2}}{2}+n'+1,
\end{equation}

From Eq.(61), (18) we reobtain the energy spectum given by the Eq.(22).

\section{Conclusion}
In this paper, we considered the generalized Kaluza-Klein monopole system. We presented the integrals of motion and the quadratic algebra for this system. We obtain for this quadratic algebra the realizations in terms of deformed oscillator algebras, the finite dimensional unitary representations and the corresponding energy spectrum. Similarly to the generalized MICZ-Kepler systems [23] and the Verrier-Evans systems studied by Daskaloyannis and  Tanoudis [55] recently, the structure constants and the Casimir operator (given by Eq.(46)) of the quadratic algbra are written in term of the Hamiltonian but other integrals. We also presented the solution using spheroidal and parabolic coordinates.

We also presented an algebraic derivation of the energy spectrum using the su(1,1) dynamical algebra obtained for the ladder operator using the radial part of the equation in spheroidal coordinates.

These results show how results obtained in context of systems without monopole on quadratic and more generally polynomial algebra can be useful for systems with monopole. 

Let us mention a new family of two-dimensional superintegrable systems with higher order integrals of motion that is a generalization of the Kepler-Coulomb system was obtained recently by Winternitz and Post [59], it would be interesting to explore the possibility of superintegrable systems with monopole allowing higher order integrals of motion. Let us mention also these recent work on superintegrable systems on curved spaces [60,61].

\ack The research of I.M. was supported by a postdoctoral
research fellowship from FQRNT of Quebec. The author thanks N.Mackay and C.Warnick for very helpful comments and discussions.

%\end{flushleft}
\end{document}